\documentclass[aps,prx,preprint,showpacs,preprintnumbers,amsmath,amssymb,floatfix,longbibliography]{revtex4-2} 

\usepackage[dvipdfmx]{graphicx}
\usepackage{bm}% bold math
\usepackage{xcolor}

\newcommand{\be}{\begin{equation}}
\newcommand{\ee}{\end{equation}}
\newcommand{\eq}[1]{Eq.~(\ref{#1})}

\def\bea{\begin{eqnarray}}
\def\eea{\end{eqnarray}}

\def\vq{{\bf q}}

\def\vk{{\bf k}}

\begin{document}

\title{Ring-like shaped charge modulations in the $t$-$J$ model with long-range Coulomb interaction} 

\author{Mat\'{\i}as Bejas$^{1}$, Roland Zeyher$^{2}$, and Andr\'es Greco$^{1,2}$}
\affiliation{
{$^1$}Facultad de Ciencias Exactas, Ingenier\'{\i}a y Agrimensura and
Instituto de F\'{\i}sica Rosario (UNR-CONICET),
Av. Pellegrini 250, 2000 Rosario, Argentina \\
{$^2$}Max-Planck-Institut f\"{u}r Festk\"{o}rperforschung, Heisenbergstra\ss{}e 1, D-70569 Stuttgart, Germany
}

\date{\today}

\begin{abstract}
The study of the charge excitations in cuprates is presently an interesting topic because of the development of new and precise x-ray experiments. Based on a large-$N$ formulation of the two-dimensional $t$-$J$ model, which allows us to consider all possible charge excitations on an equal footing, we investigate the charge spectrum for both electron- and hole-doped cases. In both cases, the instability toward phase separation, which has momentum modulation $\vq=(0,0)$, is found to be robust in a large region of the doping-temperature phase diagram. If a short-range Coulomb repulsion is included the phase separation region shrinks, but the instability remains at $\vq=(0,0)$. If on the other hand a two-dimensional long-range Coulomb interaction is included the instability sets in at $\vq$ momenta forming a ring around $\vq=(0,0)$. The computed charge spectrum in the translation-invariant phase shows well-formed rings. We discuss our results in the light of recent x-ray experiments in electron- and hole-doped cuprates, where ring-like shaped charge modulations have been reported.
\end{abstract}

\maketitle
\section{Introduction}

Recent progress in x-ray scattering techniques 
revealed the presence of a charge order (CO), with CO momentum along the axial direction [$(0,0)$-$(\pi,0)$],  in hole-doped cuprates (h-cuprates) 
\cite{ghiringhelli12,chang12,achkar12,blackburn13,blanco-canosa14,comin14,comin15,da-silva-neto14,tabis14,gerber15,tabis17} 
as well as in electron-doped cuprates  (e-cuprates) \cite{da-silva-neto15,da-silva-neto16,da-silva-neto18}.  
The CO tendency in h- and e-cuprates was intensively studied 
theoretically \cite{bejas12,allais14,meier14,wang14,atkinson15,yamakawa15,mishra15,zeyher18,sachdev13,bejas12,yamase18,yamase15,
banerjee20,caprara17,caprara17b,gannot19,nie15}, 
but its origin is still under debate. This topic is of large interest because it reveals the importance 
of charge fluctuations even in the deep underdoped region, were magnetic fluctuations are usually considered to be the dominant excitations. 

Another interesting topic of the CO is its internal symmetry. For a long time it was generally believed that $d$-wave symmetry is realized \cite{comin15}. However, recently it was found that the CO has a predominant $s$-wave symmetry \cite{mcmahon20}. More recently,  it also was found  that the  CO extends along  all directions in the same manner through the Brillouin zone (BZ), forming a ring.
Such a ring had been observed in resonant x-ray scattering in the h-cuprate Bi$_2$Sr$_2$CaCu$_2$O$_{8+\delta}$\cite{boschini21}. A similar ring has  also been reported in e-cuprates \cite{kang19}.  

In the following we present a theory for ring formation based on the $t$-$J$ model treated in the large-$N$ approximation. 
The large-$N$ approximation, where the spin index is extended from two projections to $N$ and the small parameter $1/N$ is taken for controlling the approximation, was extensively studied in Refs. [\onlinecite{morse91,zeyher96a,foussats04,greco06,wang92,affleck88a,kotliar88,grilli91,becca96,castellani95}], to name a few.
This approximation has the disadvantage that spin fluctuations are suppressed; i.e., the Mott insulator is not captured properly. However, all possible charge fluctuations can be treated on an equal footing making the method of potential interest for studying the charge excitations detected in  the new x-ray scattering experiments.
Based on the $t$-$J$ model on a square lattice and in the presence of a long-range Coulomb interaction 
($t$-$J$-$V$ model) we will show that ring-like shaped charge density waves (CDWs) are expected from the local charge susceptibility. We will discuss similarities and differences with the experiments, for both, e- and h-cuprates. 

In Sec. II we will show a summary of the theoretical method, Sec. III contains our results, Sec. IV a comparison with experiment, and Sec. V the conclusion.

\section{Mathematical framework}

We study the $t$-$J$ model on a square lattice including the two-dimensional long-range Coulomb interaction. The Hamiltonian is given by 
\begin{equation}
H = -\sum_{i, j,\sigma} t_{i j}\tilde{c}^\dag_{i\sigma}\tilde{c}_{j\sigma} + 
J \sum_{\langle i,j \rangle} \left( \vec{S}_i \cdot \vec{S}_j - \frac{1}{4} n_i n_j \right)
+\frac{1}{2} \sum_{i,j} V_{ij} n_i n_j \,,
\label{tJV}  
\end{equation}
where $\tilde{c}^\dag_{i\sigma}$ ($\tilde{c}_{i\sigma}$) are 
creation (annihilation) operators for electrons with spin $\sigma (=\uparrow, \downarrow)$  
in the Fock space without double occupancy at any site,  
$n_i=\sum_{\sigma} \tilde{c}^\dag_{i\sigma}\tilde{c}_{i\sigma}$ 
is the electron density operator, $\vec{S}_i$ is the spin operator, and 
the sites $i$ and $j$ run over a two-dimensional square lattice. 
The hopping $t_{i j}$ takes the value $t$ $(t')$ between the first (second) nearest-neighbor 
sites on the square lattice.  $\langle i,j \rangle$ denotes the nearest-neighbor sites and $J$ is the exchange interaction.
$V_{ij}$ is the long-range Coulomb interaction on the two-dimensional lattice. 

In this paper we will follow the large-N formulation of Ref.\cite{greco06}. In the large-$N$ limit
the $t$-$J$ model becomes equivalent to the following 
effective Hamiltonian in terms of usual creation and annihilation 
operators,
\begin{equation}
H_{{\rm eff}} = \sum_{{\bf k}\sigma} \epsilon({\bf k}) c^\dagger_\sigma({\bf k})
c_\sigma({\bf k}) - \frac{N_s}{2}\sum_{\alpha=1}^6 \sum_{\bf q}
\rho_\alpha^{\prime}({\bf q})\rho_\alpha^\dagger({\bf q}).
\label{Heff}
\end{equation}

At leading order, the electron dispersion $\epsilon({\vk})$ is obtained as 
\be
\epsilon({\vk})= -2 \left( t \frac{\delta}{2}+\Delta \right) (\cos k_{x}+\cos k_{y})-
4t' \frac{\delta}{2} \cos k_{x} \cos k_{y} - \mu \,.\\
\label{Epara}
\ee
\noindent While the dispersion is similar to the non-interacting case, the hopping integrals $t$ and $t'$
are renormalized by a factor $\delta/2$ where $\delta$ is the doping rate. 
The quantity $\Delta$ in \eq{Epara} is given by 
\bea{\label {Delta}}
\Delta = \frac{J}{4N_s} \sum_{\vk} (\cos k_x + \cos k_y) n_F(\epsilon(\vk)) \; , 
\eea
where $n_F$ is the Fermi function, and $N_s$ the total 
number of lattice sites.
For a given doping $\delta$, 
the chemical potential $\mu$ and $\Delta$ are determined self-consistently by solving \eq{Delta} and 
\be
(1-\delta)=\frac{2}{N_s} \sum_{\vk} n_F(\epsilon(\vk))\,.
\ee

The second term in Eq.(\ref{Heff})
represents an effective interaction, where
\begin{equation}
\rho^{\prime}_\alpha({\bf q}) = \frac{1}{N_s} \sum_{{\bf k}\sigma}
E_\alpha({\bf k},{\bf q}) c^\dagger_\sigma({\bf k}+{\bf q})c_\sigma({\bf k}),
\label{EF1}
\end{equation}
\begin{equation}
\rho_\beta({\bf q}) = \frac{1}{N_s} \sum_{{\bf k}\sigma}
F_\beta({\bf k})c^\dagger_\sigma({\bf k}+{\bf q}) c_\sigma({\bf k}),
\label{EF2}
\end{equation}
with
\begin{equation}
E_\alpha({\bf k},{\bf q}) = \left[1,t({\bf k}+{\bf q})+J({\bf q})-V(\vq),
\gamma_3({\bf k}),\gamma_4({\bf k}),\gamma_5({\bf k}),\gamma_6({\bf k}) \right],
\label{E}
\end{equation}
and
\begin{equation}
F_\beta({\bf k}) = \left[ t({\bf k}),1,2J\gamma_3({\bf k}),
2J\gamma_4({\bf k}),2J\gamma_5({\bf k}),2J\gamma_6({\bf k}) \right].
\label{F}
\end{equation}

Where $t({\bf k}) = -2 t (\cos k_{x}+\cos k_{y}) - 4t' \cos k_{x} \cos k_{y}$ and
$J(\vq) = J (\cos q_{x}+\cos q_{y})$ are the hopping and the magnetic exchange in the momentum space. $V(\vq)$ is the expression for the two-dimensional long-range Coulomb interaction in the momentum space. In Ref.\cite{becca96} the long-range Coulomb interaction $V_{ij}$ in a square plane embedded in a three-dimensional lattice is written in momentum space as  
\be
V(\vq)=\frac{V_c}{\sqrt{A^2(\vq) - 1}} \,,
\label{LRC}
\ee
where $A({\bf q})= \alpha (2 - \cos q_x - \cos q_y)+1$.
In Eq.(\ref{LRC}) $V_c= e^2 c(2 \epsilon_{\perp} a^2)^{-1}$ and 
$\alpha=\tilde{\epsilon}(a/c)^{-2}$, with  $\tilde{\epsilon}=\epsilon_\parallel/\epsilon_\perp$, where $\epsilon_\parallel$ and $\epsilon_\perp$ are the 
dielectric constants parallel and perpendicular to the planes, respectively, $a$ is the lattice constant in the plane, and $c$ the distance between them. The electric charge of electrons is denoted as $e$.

In addition, 
\begin{eqnarray}
\gamma_{3,5}& =& (\cos{k_x} \pm \cos{k_y})/2, \nonumber \\
\gamma_{4,6}& =& (\sin{k_x} \pm \sin{k_y})/2,
\label{basis}
\end{eqnarray}
where the subscripts 3 and 4 refer to the $+$, and 5 and 6 to the $-$ sign.

Using Eqs.(\ref{EF1}) and (\ref{EF2}) one easily verifies that the sum of the 
terms 
$\alpha=3$-$6$ represents the Heisenberg exchange interaction. The terms
$\alpha=1,2$ originate from the constraint which acts in $H_{\rm eff}$ as a
two-particle interaction. $H_{\rm eff}$ represents the original $t$-$J$ model in the large-$N$ limit.

In this framework, all charge susceptibilities are treated on equal footing and can be calculated as:
\begin{equation}
\chi_{\alpha\beta}(q) = \sum_{\gamma} 
\chi^{(0)}_{\alpha\gamma}(q)
[1+\chi^{(0)}(q)]^{-1}_{\gamma\beta}.
\label{chi}
\end{equation}
Here we used the abbreviation $q=(i\omega_n,{\bf q})$ with the bosonic
Matsubara frequencies $\omega_n = 2\pi T n$, where $T$ is the temperature.
$\chi^{(0)}(q)$ stands for a single bubble and is given analytically
by the expression
\begin{eqnarray}
& &\chi_{\alpha \beta}^{(0)}({\bf q},i\omega_n) = \frac{1}{N_s} \sum_{\bf k}
E_\alpha ({\bf k},{\bf q})F_\beta ({\bf k})
\frac{n_F(\epsilon({\bf k}+{\bf q}))-n_F(\epsilon({\bf k}))}
{\epsilon({\bf k}+{\bf q}) -\epsilon({\bf k})-i\omega_n}.
\label{chi00}
\end{eqnarray}

In Eq.(\ref{chi}) there are two types of charge fluctuations: on-site ($s$-wave) usual 
charge-charge correlation function, which is contained in the channels $1,2$, and bond-charge fluctuations triggered by $J$ with internal symmetry $s$ and $d$, included the flux phase \cite{cappelluti99,bejas12}, that belong to the channels $3$ to $6$. It was demonstrated in Ref.\cite{bejas17} that there is a dual structure in energy, and the channels $3$ to $6$ are practically decoupled from the channels $1,2$.
As discussed in Ref.\cite{greco06}, the usual charge-charge correlation function is
$\chi_{\rm ch}(\vq,\omega)=\chi_{12}(\vq,\omega)$ in Eq.(\ref{chi}). 
Here, we will mainly focus on $\chi_{12}(\vq,\omega)$, since the bond-charge fluctuations
play a minor role in the present study.

In the following we will consider $V_c$ and $\alpha$ in $V(\vq)$ as independent parameters. $t$ is considered the unit of energy, and $a$ the length unit. For a clear presentation of the ring structures we chose $\alpha=1$ and $V_c=1$, while in Sec. IV we use realistic values of $V_c$ and $\alpha$ in order to make contact with experiments.

It is important to note that although Eq.(\ref{chi}) looks like a usual RPA form, this expression is obtained in a large-$N$ approximation\cite{greco06}. For instance, the single bubble matrix [Eq.(\ref{chi00})] contains the form factors $E_\alpha$ and $F_\beta$ which are derived from the $t$-$J$ model in the large-N approximation.
Some results in present paper were also obtained with the equivalent formulation of Ref.\cite{bejas12}.

\section{Results}

\subsection{Results for h-cuprates}

For h-cuprates we chose $J=0.3$ and $t'=-0.3$. 

\begin{figure}
\centering
\setlength{\unitlength}{1cm}
\includegraphics[width=9cm,angle=0]{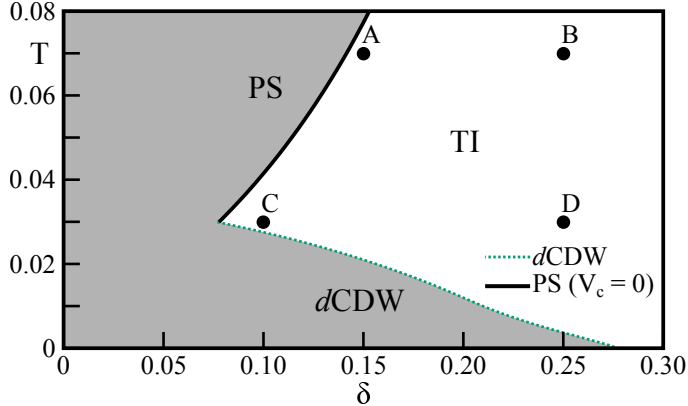}
\caption{(Color online) Phase diagram in the $\delta$-$T$ plane without Coulomb interaction, i.e., $V_c=0$. Solid (dotted) line is the onset of the phase separation ($d$CDW) instability. A, B, C, and D are four selected points in the translation-invariant phase (TI) which will be analyzed in Fig.\ref{fig:ReXhV0}.}
\label{fig:PDhV0}
\end{figure}

In Fig.\ref{fig:PDhV0} we present the phase diagram in the doping $\delta$ and temperature $T$ plane for $V_c=0$, i.e., no long-range Coulomb interaction is present. The dotted line marks the onset of the flux phase instability\cite{cappelluti99,bejas12}. The flux phase is equivalent to the
$d$-wave charge density wave ($d$CDW)
introduced phenomenologically in Ref.\cite{chakravarty01}.
The solid line represents the onset for the instability toward phase separation (PS)\cite{bejas12}.  
That is, for dopings and temperatures along the solid line the real part of the charge-charge susceptibility ${\rm Re} \chi_{\rm ch}(\vq=0,\omega=0)$ diverges.
The four points A, B, C, and D in Fig.\ref{fig:PDhV0} will be discussed below. They are located in the translation-invariant phase where our large-$N$ Fermi liquid phase is stable. In the gray region the translation-invariant phase is unstable. Our calculations are performed always in the translation-invariant phase.

\begin{figure}
\centering
\setlength{\unitlength}{1cm}
\includegraphics[width=16.1cm,angle=0]{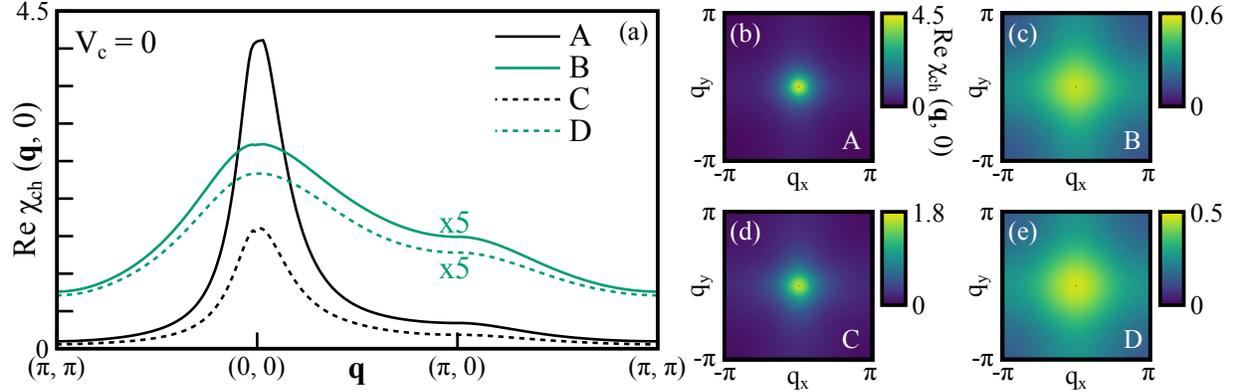}
\caption{(Color online) (a) ${\rm Re} \chi_{\rm ch}(\vq,\omega=0)$ for $\vq$ along the path $(\pi,\pi)$-$(0,0)$-$(\pi,0)$-$(\pi,\pi)$ for the selected points A, B, C, and D in Fig.\ref{fig:PDhV0}.
Note that the intensity at B and D is multiplied by $5$.
(b)-(e) Color intensity maps on the Brillouin zone of ${\rm Re} \chi_{\rm ch}(\vq,\omega=0)$ for the points A, B, C, and D, respectively.}
\label{fig:ReXhV0}
\end{figure}

\begin{figure}
\centering
\setlength{\unitlength}{1cm}
\includegraphics[width=9cm,angle=0]{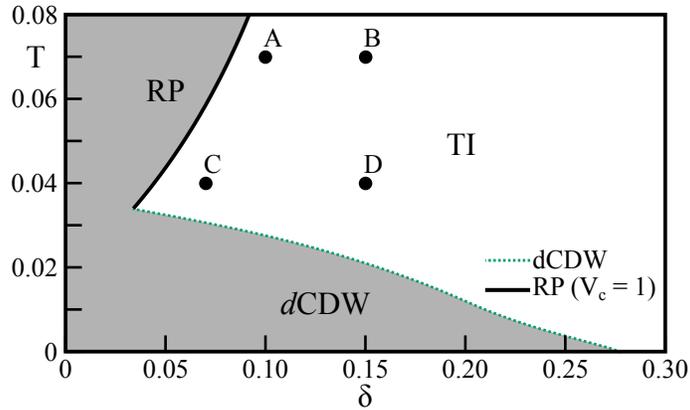}
\caption{(Color online) Phase diagram in the $\delta$-$T$ plane with long-range Coulomb interaction. Solid (dotted) line is the onset of the ring phase ($d$CDW) instability. A, B, C, and D are four selected points in the translation-invariant phase which will be analyzed in Fig. \ref{fig:ReXhV1}.}
\label{fig:PDhV1}
\end{figure}

\begin{figure}
\centering
\setlength{\unitlength}{1cm}
\includegraphics[width=16.1cm,angle=0]{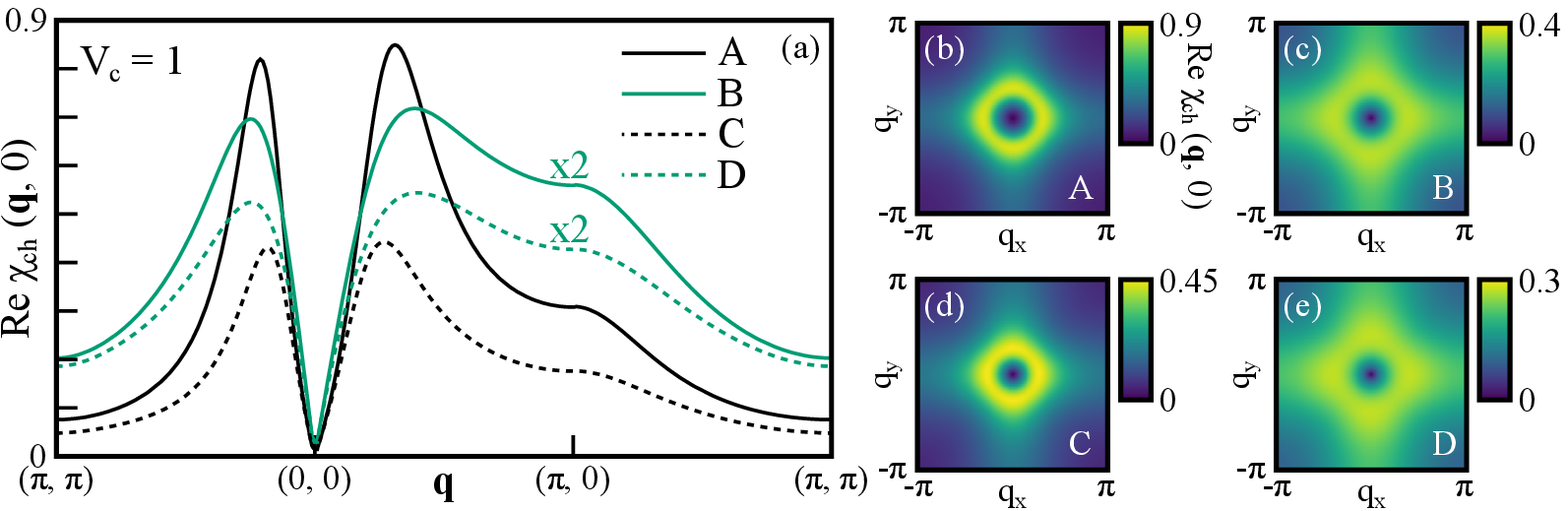}
\caption{(Color online) ${\rm Re} \chi_{\rm ch}(\vq,\omega=0)$ for $\vq$ along the path $(\pi,\pi)$-$(0,0)$-$(\pi,0)$-$(\pi,\pi)$ for the selected points A, B, C, and D in Fig.\ref{fig:PDhV1}.
(b)-(e) Color intensity maps on the Brillouin zone of ${\rm Re} \chi_{\rm ch}(\vq,\omega=0)$ for the points A, B, C, and D, respectively.}
\label{fig:ReXhV1}
\end{figure}

In Fig.\ref{fig:ReXhV0}(a) we show ${\rm Re} \chi_{\rm ch}(\vq,\omega=0)$ for $\vq$ along the path $(\pi,\pi)$-$(0,0)$-$(\pi,0)$-$(\pi,\pi)$ for the points A-D in Fig.\ref{fig:PDhV0}.
Since A and C are close to the PS instability line, ${\rm Re} \chi_{\rm ch}(\vq,\omega=0)$ shows sharp and large peaks located at $\vq=(0,0)$. If both A and C move to lower dopings and touch the solid line they will show a $\delta$ function at $\vq=(0,0)$ signaling the long-range order for PS.
The points B and D also show peaks at $\vq=(0,0)$, but they are broader and less intense than in A and C. That is because the points B and D are far from the PS line, and the fluctuations of the PS phase are weaker in B and D than in A and C. The point C is also close to the $d$CDW instability line; however, as discussed above, bond-charge and the usual charge-charge correlations are decoupled. Then, the effect of $d$CDW on $\chi_{\rm ch}$ is negligible.
The peaks at ${\bf q}=(0,0)$ appear as disks in the color intensity maps for ${\rm Re}\chi_{\rm ch}(\vq,\omega=0)$ over the full BZ depicted in Fig.\ref{fig:ReXhV0}(b)-(e). For the points
A and C, close to the PS instability line, the disks are small and intense while for B and D are broader and with much lower intensity.

The situation changes when we include the long-range Coulomb repulsion. In Fig.\ref{fig:PDhV1} we show the phase diagram including the long-range Coulomb interaction $V(\vq)$. Note that the $d$CDW is not affected by the presence of the long-range Coulomb interaction, showing the decoupling between channels $3$-$6$ and the channels $1,2$ as discussed above. The solid line is clearly shifted to lower dopings showing a larger region in the phase diagram where the translation-invariant phase is stable. However, the nature of the charge instability at the solid line line is no longer PS.
In Fig.{\ref{fig:ReXhV1}}(a) we show the ${\rm Re} \chi_{\rm ch}(\vq,\omega=0)$ for $\vq$ along the path $(\pi,\pi)$-$(0,0)$-$(\pi,0)$-$(\pi,\pi)$ for the points located at A-D in Fig.\ref{fig:PDhV1}. Now, in the presence of the long-range Coulomb interaction, the peaks are not located at $\vq=(0,0)$ but instead they are shifted from $(0,0)$. As in the case of Fig.\ref{fig:ReXhV0}, the peaks are more intense and sharper for A and C than for B and D. 
Figures \ref{fig:ReXhV1}(b)-(e) show color intensity maps for ${\rm Re}\chi_{\rm ch}(\vq,\omega=0)$ on the BZ for the selected points A, B, C, and D on the phase diagram, respectively. Interestingly, the intensity maps show clear and well-defined ring-shaped charge modulations around $\vq=(0,0)$. For this reason we have called the phase that occurs at the onset of the solid line the ring phase (RP).
Note that, as they are farther from the instability line, the rings for B and D are less intense and more diffused than for A and C.

\subsection{Results for e-cuprates}

The $t$-$J$ model is defined in the space where the presence of two electrons per site is prohibited. In order to present calculations for e-cuprates we proceed as usual: we make a particle-hole transformation which requires a sign change for $t'$\cite{martins01,tohyama94,tohyama99,bejas14}.
Here we chose $J=0.3$ and $t'=0.3$.

\begin{figure}
\centering
\setlength{\unitlength}{1cm}
\includegraphics[width=16.0cm,angle=0]{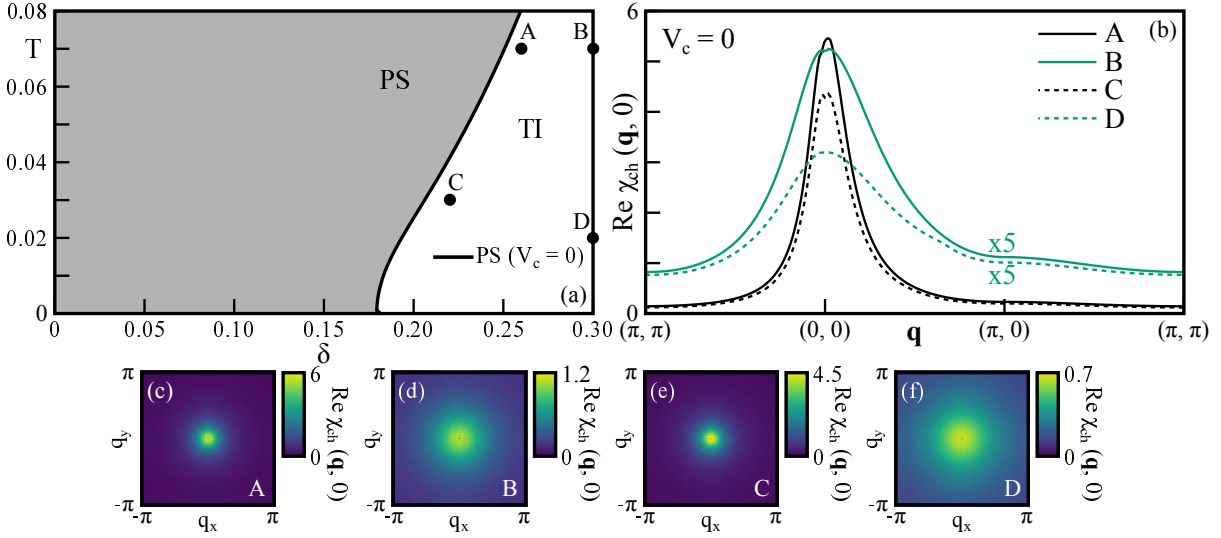}
\caption{(Color online) (a) Phase diagram for e-cuprates without Coulomb interaction.
Momentum cuts (b), and maps (c)-(f) of ${\rm Re}\chi_{\rm ch}(\vq,\omega=0)$ for the selected points A-D.}
\label{fig:eV0}
\end{figure}

\begin{figure}
\centering
\setlength{\unitlength}{1cm}
\includegraphics[width=16.0cm,angle=0]{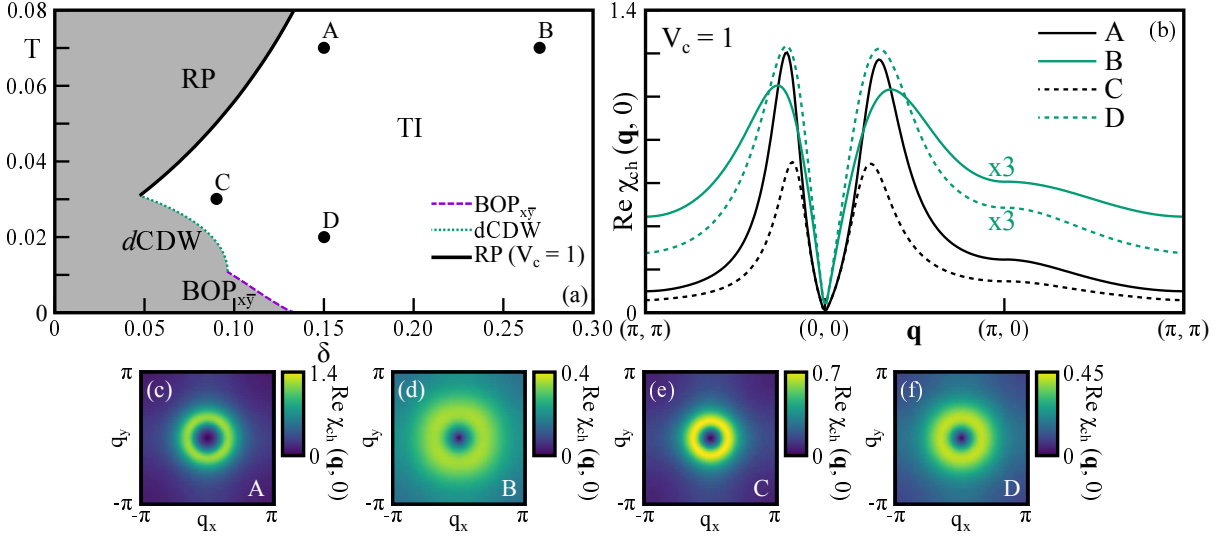}
\caption{(Color online) (a) Phase diagram for e-cuprates with long-range Coulomb interaction. In the presence of a long-range Coulomb interaction the homogeneous translation-invariant phase extends and $d$CDW (dotted line) and BOP$_{x\bar{y}}$ instabilities occur.
Momentum cuts (b), and maps (c)-(f) of ${\rm Re}\chi_{\rm ch}(\vq,\omega=0)$ for the selected points A-D.}
\label{fig:eV1}
\end{figure}

Figure \ref{fig:eV0}(a) shows the phase diagram for $V_c=0$. Comparing Fig.\ref{fig:PDhV0} and Fig.\ref{fig:eV0}(a) we can see that there is a larger tendency for PS in the electron-doped case\cite{bejas14}. 
Figure \ref{fig:eV0}(b) is equivalent to Fig.\ref{fig:ReXhV0} for the h-cuprates, and show that for $V_c=0$ the local charge instability is toward the usual PS with ordered momentum $\vq=0$.
Then, the momentum map shows disks [Fig.\ref{fig:eV0}(c)-(f)] instead of rings.
Similar to Fig.\ref{fig:PDhV1} and Fig.\ref{fig:ReXhV1}, Fig.\ref{fig:eV1} shows that rings are formed in the translation-invariant phase near the RP instability when the long-range Coulomb is included.
That is, ring-shaped charge modulations also occurs in e-cuprates.
Since now the region for the translation-invariant phase is much more extended in Fig.\ref{fig:eV1}(a) than in Fig.\ref{fig:eV0}(a), we also show the leading instabilities from the channels $3$-$6$, triggered by $J$. The flux phase is the leading phase in the temperature range $0.01$-$0.03$ and $\delta < 0.10$. For larger doping and lower temperatures 
a bond-order phase (BOP$_{x\bar{y}}$)
is the leading instability\cite{bejas14}.

\section{Comparison with experiments}

Resonant x-ray scattering on Bi$_2$Sr$_2$CaCu$_2$O$_{8+\delta}$ discovered a dynamic quasicircular charge pattern \cite{boschini21}, i.e, ring-like shaped charge modulations. The authors of Ref.\cite{boschini21} proposed an explanation for the rings.
Assuming the random phase approximation (RPA)\cite{mahan} and a Coulomb repulsion of the form $V(\vq)=V_c(\vq)+U(\vq)$, where $V_c(\vq)$ is a two-dimensional long-range Coulomb interaction and $U(\vq)$ is a short-range Coulomb potential, the charge-charge correlation function was calculated. Although both $V_c(\vq)$ and $U(\vq)$ 
have a monotonic behavior, the sum of both show a nonmonotonic behavior with a minimum at a given $\vq$ that determines the quasicircular pattern.
Besides the proposed RPA does not describe the observed rings in detail, RPA works for weakly correlated systems but cuprates are considered strongly correlated.

\begin{figure}
\centering
\setlength{\unitlength}{1cm}
\includegraphics[width=8.9cm,angle=0]{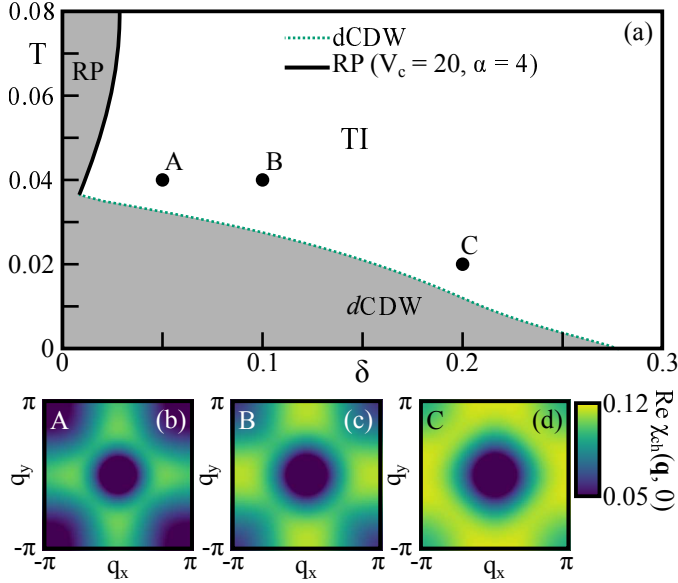}
\caption{(Color online) (a) Phase diagram for h-cuprates for realistic long-range Coulomb parameters.
(b)-(d) Color intensity map on the Brillouin zone for ${\rm Re} \chi_{\rm ch}(\vq,\omega=0)$ for the selected points A, B, and C, respectively.}
\label{fig:V20}
\end{figure}
In this paper we have studied the $t$-$J$ model, which is a model more appropriated for cuprates\cite{fczhang88}. The $t$-$J$ model contains the competition between 
kinetic energy and PS\cite{bejas14,martins01,tohyama94,macridin06}, which is characterized by a strong peak at $\vq=(0,0)$ in the usual charge-charge correlation function. This competition originates in the constraint that prohibits double occupancy, and it is missing in RPA.
It is well known that the inclusion of a short-range Coulomb repulsion reduces the PS region in the doping and temperature phase diagram\cite{bejas14}. However, if instead of including a short-range we applied a long-range Coulomb interaction, we found rings as shown in the previous sections. Interestingly the radii of the rings are of the order of $0.3\pi$-$0.4\pi$, which is close to the radius reported in the experiment\cite{boschini21,kang19}; see for instance Fig.\ref{fig:ReXhV1}(b)-(e).

In previous sections, for a clearer presentation of the results we have chosen the parameter values $\alpha=1$ and $V_c=1$ in the long-range Coulomb potential.
What happens if we use more suitable parameters as those used for describing the recently observed low-energy plasmons in resonant inelastic x-ray scattering (RIXS)\cite{greco20,nag20,hepting22}? In Fig.\ref{fig:V20} we present results for h-cuprates ($t'=-0.3$), and we chose $\alpha=4$ and $V_c=20$ which are of the order of those used in Refs.\cite{greco20,nag20,hepting22} in the context of the $t$-$J$-$V$ model. The instability line toward the ring phase is much more shrunken.

In Figs.\ref{fig:V20}(b)-(d) we show color intensity maps for the three chosen points A, B, and C in panel (a). These intensity maps show that the largest intensity occurs along the axial direction as in the experiment\cite{boschini21}. Then, in contrast to the theoretical study presented in Ref.\cite{boschini21}, our calculation captures the 
anisotropy behavior of the quasicircular charge pattern. In addition, as discussed in Ref.[\onlinecite{boschini21}], our rings are dynamic, since they occur in the translation-invariant phase.

Up to now we have discussed the ring-shaped structures by analyzing 
${\rm Re}\chi_{\rm ch}({\bf q}, \omega = 0)$. In Ref.[\onlinecite{boschini21}] the rings where discussed by using resonant x-ray scattering (RXS) and energy-integrated RIXS. 
RXS measures the equal-time correlation function,
\begin{align}
I_{RXS}(\bf{q}) &= \int_{-\infty}^{\infty} \rm{Im}\chi_{ch}({\bf q},\omega) [n_B(\omega)+1] d\omega \nonumber\\
&= \int_{0}^{\infty} \rm{Im}\chi_{ch}({\bf q},\omega) [2n_B(\omega)+1] d\omega
\label{eq:IRXS}
\end{align}
\noindent where $n_B(\omega)$ is the Bose factor, 
and ${\rm Im} \chi_{ch}({\bf q},\omega)$ is proportional to the RIXS spectrum. 

\begin{figure}
\centering
\setlength{\unitlength}{1cm}
\includegraphics[width=8.9cm,angle=0]{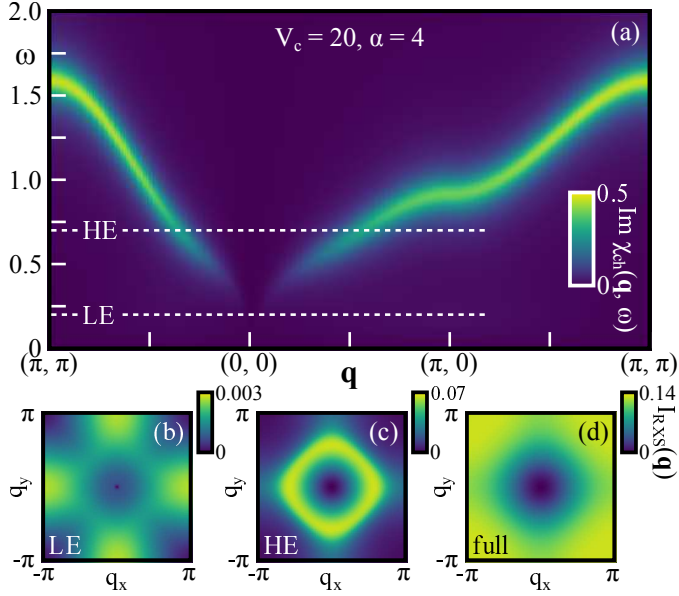}
\caption{(Color online) (a) Momentum-energy map of ${\rm Im} \chi_{\rm ch}(\vq,\omega)$ along the path $(\pi,\pi)$-$(0,0)$-$(\pi,0)$-$(\pi,\pi)$ for the point B in Fig.\ref{fig:V20}(a). The horizontal dashed lines indicate the limits for the energy integration discussed in the text.
(b) Color intensity map on the Brillouin zone of the low-energy (LE) integration of ${\rm Im} \chi_{\rm ch}(\vq,\omega)$.
(c) The same as (b) for the high-energy (HE) integration.
(d) The same as (b) and (c) for the full energy range, i.e., from $\omega=0$ to infinity.}
\label{fig:ImX}
\end{figure}

In Fig.\ref{fig:ImX}(a) we show the momentum-energy map of ${\rm Im} \chi_{\rm ch}(\vq,\omega)$ for the point B in Fig.\ref{fig:V20}(a) along the path $(\pi,\pi)$-$(0,0)$-$(\pi,0)$-$(\pi,\pi)$. We obtain a two-dimensional plasmon branch which has similarities to the low-energy plasmons observed in cuprates\cite{hepting18,hepting22,nag20,lin20,singh22}. 
What happens if we integrate the spectra of ${\rm Im} \chi_{\rm ch}({\bf q, \omega})$ [Eq.(\ref{eq:IRXS})] in different energy windows?
Figure \ref{fig:ImX}(b) shows the result for a low-energy (LE) integration up to the dashed line indicated with LE in Fig.\ref{fig:ImX}(a). As in the experiment, Fig.\ref{fig:ImX}(b) shows
the largest intensity along the axial directions.
The obtained map is similar to that of Fig.\ref{fig:V20}(c) for $\rm{Re}\chi_{ch}({\bf q}, \omega = 0)$, since the low energy integration in Eq.(\ref{eq:IRXS}) leads to $I_{RXS}(\bf{q}) \sim \rm{Re}\chi_{ch}({\bf q}, \omega = 0)$. If instead we integrate in a high-energy window, up to the dashed line indicated with HE in Fig.\ref{fig:ImX}(a), we collect contributions from the V-shaped dispersion, and rings are formed [Fig.\ref{fig:ImX}(c)]. However, the signals along the axial and nodal direction have similar intensity; the stronger intensity along the axial direction is lost.
If we integrate the entire energy spectra the rings are washed out [Fig.\ref{fig:ImX}(d)], due to the contribution from the top of the 2D plasmon dispersion.

Although our model does not capture quantitatively all the experimental aspects, as the doping and temperature behavior of the intensity and radius of the rings, we think that the similarities between our results and those reported in Ref.\cite{boschini21} for Bi$_2$Sr$_2$CaCu$_2$O$_{8+\delta}$ are remarkable. In fact, we agree with the authors of Ref.\cite{boschini21} on the importance of the long-range Coulomb interaction for the formation of the rings.
It would be desirable to make more x-ray experiments for different members of the cuprate family as a function of doping and temperature.

Finally, in Ref.[\onlinecite{boschini21}] it is suggested that the momentum of the observed stronger intensity along the axial direction matches the wave vector of the discussed axial CO\cite{ghiringhelli12,chang12,achkar12,blackburn13,blanco-canosa14,comin14,comin15,da-silva-neto14,tabis14,gerber15,tabis17}.
It is important to mention that most of the theoretical studies about the CO   \cite{bejas12,allais14,meier14,wang14,atkinson15,yamakawa15,mishra15,zeyher18,sachdev13,bejas12,yamase18,yamase15,banerjee20} suggest that the CO occurs due to bond-order charge excitations, and have an internal form factor different from the isotropic $s$ wave. The rings appear in our calculation in the usual $s$-wave charge-charge correlation function, as also proposed in Ref.[\onlinecite{boschini21}]. While bond-order charge fluctuations occur in our calculation in the sector $3$-$6$, usual charge excitations occur in the $1$-$2$ sector, and both channels are mainly  decoupled\cite{bejas17}. This is an interesting topic which needs more discussion, because a bond-order charge fluctuation scenario is completely different from the usual $s$-wave charge fluctuation picture.

\section{Conclusion}

Phase separation is an instability at $\vq=(0,0)$ of the  charge-charge correlation function. PS exists in the $t$-$J$-$V$ model even for $J=V=0$.
Thus, only the first term in Eq.(\ref{tJV}) is present.
This means that the constraint which prohibits 
double occupancy plays a central role in the instability toward PS.  
It is well known that PS may cover a large region of the phase diagram, and  the inclusion of a short-range Coulomb interaction reduces the PS region.   
This picture changes drastically if a long-range Coulomb interaction is included to cancel PS.
In this case, although the instability region is reduced, the momenta $\vq$ where the instability takes place is no longer $\vq=(0,0)$ and the ordered momenta follows a {\bf ring-like} shape around $\vq=(0,0)$, leading to what we have called a ring phase. Interestingly, the charge excitation spectrum in the translation-invariant phase shows well-defined ring-shaped charge modulations, for both e-cuprates and h-cuprates.
It is worth pointing out that although we compute the charge-charge correlation function at $\omega = 0$, this does not mean that the rings are static.
As we are in the translation-invariant phase, the rings are formed by fluctuations in the proximity of the ring phase instability.
The presence of rings is not difficult to understand.
The long-range Coulomb interaction is strongly peaked at $\vq=(0,0)$ and decays with increasing $\vq$.
If the system shows PS, under the presence of the long-range Coulomb interaction the charge accumulation at $\vq=(0,0)$ is expelled, and rings are developed.
Thus, the wave vector ${\bf Q}$, whose magnitude is associated with the radius of the ring, is unrelated to the Fermi surface topology and depends on the presence of the long-range Coulomb interaction.
${\bf Q}$ is defined as the momentum ${\bf q}$ where the maximum of ${\rm Re} \chi_{\rm ch}(\vq,\omega=0)$ is located.
To quantify the doping dependence, for instance, Fig.\ref{fig:ReXhV1} (a) shows that while the doping almost doubles between C and D, $|{\bf Q}|$ increases by $\sim 50\%$ in the axial direction and by $\sim 35\%$ in the diagonal direction.
In addition, as can be seen also in Figs.\ref{fig:ReXhV1} (b)-(e) and Figs.\ref{fig:eV1} (c)-(f), the intensity of the ring decreases and the ring widens with increasing doping away from the RP instability line.
It is important to mention that the ring-shaped structures are not exactly rings in the mathematical sense.
Although within the thickness of the RP instability line the static charge susceptibility diverges for all wave vectors along the ring, which motivates the name RP, in the translation-invariant phase the intensity of the ring has a modulation with a maximum at the axial (diagonal) direction for h-cuprates (e-cuprates).
In Figs.\ref{fig:ReXhV1} (b)-(e) (Figs.\ref{fig:eV1} (c)-(f)) we can see well-defined rings; however they have a very weak intensity modulation, indiscernible in the intensity maps but noticeable in the cuts in Fig.\ref{fig:ReXhV1} (a) (Fig.\ref{fig:eV1} (b)), with maximum at the axial (diagonal) direction.

$|{\bf Q}|$ depends also on the strength of the long-range Coulomb interaction $V_c$.
$|{\bf Q}|$ increases with increasing $V_c$ which can be seen after comparing Figs.\ref{fig:ReXhV1} (b)-(e) and Figs.\ref{fig:eV1} (c)-(f), for $V_c = 1$, with 
Figs.\ref{fig:V20} (b)-(d) for the realistic value of $V_c=20$.
In addition, the  anisotropy along the ring is more evident at low energy [Fig.\ref{fig:V20} (b) and Fig. \ref{fig:ImX} (b)] for realistic values of the long-range Coulomb interaction, where, as in the experiment, the largest intensity is at the axial directions.

Our results capture several important details of the experiment. Dynamic ring-shaped charge modulations exist in the presence of the long-range Coulomb interaction in the translation-invariant phase. However, it is a prerequisite for the existence of rings that without long-range $V(\vq)$ the system shows PS. 

\acknowledgments
We thank M. Hepting, A.M. Ole\'s, and H. Yamase for fruitful discussions, and M. Minola and E. H. da Silva Neto for a critical reading of the manuscript. A.~G. acknowledges the Max Planck Institute for Solid State Research in Stuttgart for hospitality and financial support.
A part of the results presented in this work was obtained by using the facilities of the CCT-Rosario Computational Center, member of the High Performance Computing National System (SNCAD, MincyT-Argentina).

\bibliography{main} 

\end{document}